\documentclass{article}
\begin{document}

\title{The Levi-Civita tensor noncovariance and curvature in the pseudotensors space}%
\author{A. L. Koshkarov\thanks{email: Koshkarov@petrsu.ru}}%
\date{University of Petrozavodsk, Physics department, Russia}
\maketitle
\begin{abstract}
It is shown that conventional "covariant" derivative of the Levi-Civita tensor
$E_{\alpha\beta\mu\nu;\xi}$ is not really covariant. Adding compensative terms,
it is possible to make it covariant and to be equal to zero.
Then  one can be introduced a curvature in the pseudotensors space.
 There appears a curvature tensor which is dissimilar to ordinary one by {\em covariant}
term including the Levi-Civita density derivatives hence to be equal to zero. This term is a little
bit
similar to Weylean one in the Weyl curvature tensor.
There has been attempted to find a curvature measure in the combined (tensor plus pseudotensor)
tensors space. Besides, there has been constructed some vector from the metric and the Levi-Civita
density which gives new opportunities in geometry.
\end{abstract}
%
\section{Introduction}

Tensor analysis which General Relativity is based on operates basically with true tensors
and it is very little spoken of pseudotensors role. Although there are many objects
in the world to be bound up with pseudotensors. For example most of particles and bodies in the
Universe have  angular momentum  or
 spin.

Here is a question: could a curvature tensor be a pseudotensor or include
that in some way? It's not clear. Anyway, symmetries of the Riemann-Christopher
tensor are not compatible with those of the Levi-Civita pseudotensor.

There are examples of using values in Physics to be a sum of a tensor and a pseudotensor.
Let's recall for example ($V-A$)-hamiltonian~\cite{fgms} in the Weak Interaction Physics. All that
leads to  parity breakdown, reflection symmetry violation and so forth.

Formally the Levi-Civita pseudotensor transforms like a true tensor, and
covariant differentiation procedure can be applied. And then it  turned
out the value $T_{\alpha\beta\mu\nu;\rho}$
("covariant" derivative of the Levi-Civita pseudotensor)
not to be covariant, i. e.  not to be a tensor.

On the base of the result, there works out   differential geometry in the
pseudotensors space (it is constructed right covariant derivative of the
Levi-Civita pseudotensor $D_\rho E_{\alpha\beta\mu\nu}$, introduced an extended
connection and a covariant derivative). The curvature tensor is calculated.

Among other things, some new vector is built from the metric and the Levi-Civita density.
That gives  new opportunities in geometry. For instant, one can extend a connection adding
new  tensor or pseudotensor associated with the vector.
Particularly, an example of a connection to be invariant under Weylean rescaling without any
gradient transformation is given. Also in more detail, the connection with the pseudotensor
extension is treated.

\section{D4-covariant derivative of the Levi-Civita tensor}

The Levi-Civita tensor  density $\varepsilon_{\alpha\beta\mu\nu}$ is
defined in the same way in each coordinate system
$$
\varepsilon_{\alpha\beta\mu\nu}=
\left\{
\begin{array}{rcl}
\pm1,\quad \alpha,\beta,\mu,\nu -\mbox{ are all different}\\
0,\qquad \qquad\mbox{ otherwise} \\
\end{array}
\right.
$$

Together with the density $\varepsilon_{\alpha\beta\mu\nu}$ we are going to
treat the Levi-Civita {\em pseudotensor} $E_{\alpha\beta\mu\nu}$
$$
E_{\alpha\beta\mu\nu}=\sqrt{-g}\,\varepsilon_{\alpha\beta\mu\nu}
$$
Formally it transforms like a true tensor, but its contraction with an arbitrary
4-rank tensor results a pseudoscalar. For this reason it is
called a pseudotensor.

Now let us show that ordinary covariant derivative (with the metric-compatible
Levi-Civita or metric or Riemannian connection) of the Levi-Civita tensor $E_{\alpha\beta\mu\nu;\xi}$
is noncovariant, i.e. nontensorial value. We'll obtain a formula
$$
E_{\alpha\beta\mu\nu;\xi}=\sqrt{-g}\,\varepsilon_{\alpha\beta\mu\nu,\xi}
$$

By definition of covariant derivative
$$
E_{\alpha\beta\mu\nu;\xi}=E_{\alpha\beta\mu\nu,\xi}-
\Gamma^\lambda_{\alpha\xi}E_{\lambda\beta\mu
\nu}-\Gamma^\lambda_{\beta\xi}E_{\alpha\lambda\mu\nu}-
\Gamma^\lambda_{\mu\xi}E_{\alpha\beta\lambda
\nu}-\Gamma^\lambda_{\nu\xi}E_{\alpha\beta\mu\lambda}
$$
Let's take here
 $\alpha,\beta,\mu,\nu=0,1,2,3$:
$$
E_{0123;\xi}=E_{0123,\xi}-\Gamma^\lambda_{0\xi}E_{\lambda123
}-\Gamma^\lambda_{1\xi}E_{0\lambda23}-\Gamma^\lambda_{2\xi}E_{01\lambda
3}-\Gamma^\lambda_{3\xi}E_{012\lambda}=
E_{0123,\xi}-\Gamma^\lambda_{\lambda\xi}E_{0123}=
$$
$$
=E_{0123,\xi}-(\frac{(\sqrt{-g})_{,\xi}}{\sqrt{-g}}E_{0123}
$$
Consequently
$$
E_{\alpha\beta\mu\nu;\xi}=E_{\alpha\beta\mu\nu,\xi}-
\frac{(\sqrt{-g})_{,\xi}}{\sqrt{-g}}E_{\alpha
\beta\mu\nu}=(\sqrt{-g}\,\varepsilon_{\alpha\beta\mu\nu})_{,\xi}-
(\sqrt{-g})_{,\xi}\,\varepsilon_{
\alpha\beta\mu\nu}=\sqrt{-g}\,\varepsilon_{\alpha\beta\mu\nu,\xi},
$$
and we have obtained a formula above.

In the right hand side we have zero in the given coordinate system since the components
$\varepsilon_{\alpha\beta\mu\nu}$ are constant.  But this 'zero'  is not covariant since
the expression in the right hand side is not a tensor. Hence, there will be nonzero
expression for $E_{\alpha\beta\mu\nu;\xi}$ in the other coordinate system.

Similarly one can be treated a "covariant" derivative of the Levi-Civita tensor density
$$
\varepsilon_{\alpha\beta\mu\nu;\xi}=\varepsilon_{\alpha\beta\mu\nu,\xi}-
\Gamma^\lambda_{\alpha\xi}
\varepsilon_{\lambda\beta\mu\nu}
-\Gamma^\lambda_{\beta\xi}\varepsilon_{\alpha\lambda\mu\nu}-
\Gamma^\lambda_{\mu\xi}
\varepsilon_{\alpha\beta\lambda\nu}-\Gamma^\lambda_{\nu\xi}\varepsilon_{
\alpha\beta\mu\lambda}+
$$
$$
+\frac{(\sqrt{-g})_{,\xi}}{\sqrt{-g}}\varepsilon_{\alpha\beta\mu\nu}=
\varepsilon_{\alpha\beta\mu\nu,\xi}
$$
We can see that a metric connection compensates the metric part of the density derivative
and does not compensate the very derivative $\varepsilon_{\alpha\beta\mu\nu,\xi}$. As a result,
 one obtained noncovariant expression  for the "covariant derivative".

It is necessary of course to mention the metric tensor determinant which is a tensor density
as well. The Levi-Civita density is involved there. So  is  the covariant derivative $g_{;\xi}$
to be equal zero really covariant? At this point, it seems to be all right. When applied
the covariant differentiation rule, one can be seen that derivative $g_{,\xi}$  is compensated by
the extra density term to contain contracted Christoffel symbol.

It is interesting to note that the contracted Christoffel symbol includes the metric determinant.
Therefore, the metric connection is still not fully independent of the Levi-Civita density.

So, it is not enough to have a metric connection only in order to construct a right covariant
derivative of the Levi-Civita tensor. One needs extra compensators which can be expressed in terms
of the Levi-Civita density derivatives.

It will be shown further, on the base of this result one can construct a right covariant
derivative $D_\xi E_{\alpha\beta\mu\nu}$. Like the metric covariant derivative, the derivative
$D_\xi E_{\alpha\beta\mu\nu}$ will be equal to zero. Then there will be determined a curvature
in the pseudotensors space.

To build a covariant derivative of the Levi-Civita tensor we use a tensorial relation
$$
E_{\alpha\beta\mu\nu}E^{\rho\sigma\gamma\delta}=
-\delta^{\rho\sigma\gamma\delta}_{\alpha\beta\mu\nu}
=-\left|\begin{array}{cccc}
  \delta_{\alpha}^{\rho} & \delta_{\beta}^{\rho}& \delta_{\mu}^{\rho} & \delta_{\nu}^{\rho} \\
  \delta_{\alpha}^{\sigma}& \delta_{\beta}^{\sigma} & \delta_{\mu}^{\sigma} & \delta_{\nu}^{\sigma}\\
  \delta_{\alpha}^{\gamma} & \delta_{\beta}^{\gamma} & \delta_{\mu}^{\gamma}& \delta_{\nu}^{\gamma} \\
  \delta_{\alpha}^{\delta} & \delta_{\beta}^{\delta} & \delta_{\mu}^{\delta} & \delta_{\nu}^{\delta}
\end{array}\right|,
$$
that covariantly differentiates
\begin{equation}\label{edelta;=0}
(E_{\alpha\beta\mu\nu}E^{\rho\sigma\gamma\delta})_{;\xi}=
E_{\alpha\beta\mu\nu;\xi}E^{\rho\sigma\gamma\delta}+
E_{\alpha\beta\mu\nu}E^{\rho\sigma\gamma\delta}{}_{;\xi}=0
\end{equation}

The left hand side of the equality is a 0-tensor, but each term is not a tensor. Therefore any
of these terms can be considered as a compensator. Multiplying both sides by
$E_{\rho\sigma\gamma\delta}$,  we obtain
\begin{equation}\label{compens}
E_{\alpha\beta\mu\nu;\xi}-
\frac{1}{24}(E_{\rho\sigma\gamma\delta}E^{\rho\sigma\gamma\delta}{}_{;\xi})
E_{\alpha\beta\mu\nu}=0
\end{equation}
Denoting
 $\gamma_\xi=E_{\rho\sigma\gamma\delta}E^{\rho\sigma\gamma\delta}{}_{;\xi}=
 \varepsilon_{\rho\sigma\gamma\delta}\varepsilon^{\rho\sigma\gamma\delta}{}_{,\xi}$,
we have
\begin{equation}\label{covarE}
D_\xi E_{\alpha\beta\mu\nu}\equiv E_{\alpha\beta\mu\nu;\xi}-\frac{1}{24}
\gamma_\xi E_{\alpha\beta\mu\nu}=0
\end{equation}
Once again should be reminded that $\gamma_\xi$ is zero in the given coordinate system.

Now we  in reality have covariant derivative of the Levi-Civita tensor that is equal to zero.
Formally this leads to connection's modification and covariant derivative for pseudotensors
and consequently to modification of the curvature tensor in the pseudotensors space.

Note the transformation rule of $\gamma_\xi$
\begin{equation}\label{preogam}
\gamma_\xi=\frac{\partial x'^{\xi_1}}{\partial x^\xi}\left(\gamma'_{\xi_1}-24
\frac{|\partial x/\partial x'|_{,\xi_1}}{|\partial x/\partial x'|}\right)
\end{equation}
Here $|\partial x/\partial x'|$ is a Jacobian. It is of interest to compare it to the contracted
Christoffel symbol transformation rule
$\Gamma_\xi\equiv\Gamma^\lambda_{\xi\lambda}=(\ln\sqrt{-g})_{,\xi}=(\sqrt{-g})_{,\xi}/\sqrt{-g}$:
\begin{equation}\label{preoGam}
\Gamma_\xi=\frac{\partial x'^{\xi_1}}{\partial x^\xi}\left(\Gamma'_{\xi_1}+
\frac{|\partial x'/\partial x|_{,\xi_1}}{|\partial x'/\partial x|}\right)=
\frac{\partial x'^{\xi_1}}{\partial x^\xi}\left(\Gamma'_{\xi_1}-
\frac{|\partial x/\partial x'|_{,\xi_1}}{|\partial x/\partial x'|}\right)
\end{equation}

Now let's form an object $G_\xi(q)=\Gamma_\xi+q\gamma_\xi$ to be transforming as follows
$$
G_\xi(q)=
\frac{\partial x'^{\xi_1}}{\partial x^\xi}\left(\Gamma'_{\xi_1}+q\gamma'_{\xi_1}-(1+24q)
\frac{|\partial x/\partial x'|_{,\xi_1}}{|\partial x/\partial x'|}\right)
$$
Choose $q=q_0=-1/24$:
\begin{equation}\label{Qpreob}
G_\xi(q_0)=\frac{\partial x'^{\xi_1}}{\partial x^\xi}G'_\xi(q_0)
\end{equation}
It means that $G_\xi(q_0)$  is a true vector!

Strictly, the vector is not a gradient but practically it is the gradient in the given coordinate system,
for  the manifestly nongradient term $\gamma_\xi$ is equal to zero.

Existence of the vector opens new opportunities.
For example, one can build new connections
\begin{equation}\label{tconnect}
\widetilde{\Gamma}_{\alpha\mu\xi}=\Gamma_{\alpha\mu\xi}+G_\alpha(q_0) g_{\mu\xi}-G_\mu(q_0)g_{\alpha\xi}
\end{equation}
or
\begin{equation}\label{pconnect}
\widetilde{\Gamma}_{\alpha\mu\xi}=\Gamma_{\alpha\mu\xi}+G^\lambda(q_0)E_{\lambda\alpha\mu\xi}
\end{equation}
These connections do not violate the metricity condition. The connection~(\ref{pconnect})
 differs from metric one by a pseudotensor
and leads to the combined (tensor+pseudotensor) curvature tensor.

Another opportunities is a connection of the Weylean type
\begin{equation}\label{wconnect}
\widetilde{\Gamma}_{\alpha\mu\xi}=\Gamma_{\alpha\mu\xi}-\frac{1}{4}
(g_{\alpha\mu}G_\xi(q)+g_{\alpha\xi}G_\mu(q)-g_{\mu\xi}G_\alpha(q))
\end{equation}
For example, the connection
\begin{equation}\label{wconnect1}
\widetilde{\Gamma}^\lambda{}_{\mu\xi}(q_0)=g^{\lambda\alpha}\{\Gamma_{\alpha\mu\xi}-\frac{1}{4}
(g_{\alpha\mu}G_\xi(q_0)+g_{\alpha\xi}G_\mu(q_0)-g_{\mu\xi}G_\alpha(q_0))\}
\end{equation}
is invariant under the Weylean rescaling $(g_{\mu\nu}\rightarrow sg_{\mu\nu})$ automatically,
since in the transformation, the vector $G_\mu(q_0)$ transforms as follows
$$
G^s_\mu(q_0)=G_\mu(q_0)+2(\ln s)_{,\mu}
$$
And no need to make any gradient transformation of the vector $G_\mu(q_0)$!

The vector $G_\mu(q_0)$ is not a gradient, so it cannot be annulled with gradient transformation.

The transformations rules~(\ref{preogam}) and~(\ref{preoGam}) put some thought to treat some
special conformal transformation with Jacobian as a dilatation parameter. Namely, the
connection~(\ref{wconnect}) can be made invariant under such a transformation.

Further the case~(\ref{pconnect}) will be treated in more details.

It is of interest another way of transcript of a formula~(\ref{compens}). Let us
multiply~(\ref{edelta;=0}) by $E_{\tau\kappa\eta\zeta}$:
$$
-(\delta^{\rho\sigma\gamma\delta}_{\tau\kappa\eta\zeta}
E_{\alpha\beta\mu\nu})_{;\xi}+
E_{\alpha\beta\mu\nu}E_{\tau\kappa\eta\zeta}E^{\rho\sigma\gamma\delta}
{}_{;\xi}=0
$$
Taking $\rho,\sigma,\gamma,\delta=\alpha,\beta,\mu,\nu$, we obtain~(\ref{compens}). If
$\rho,\sigma,\gamma,\delta=\beta,\mu,\nu,\xi$, then
$$
-6E_{\tau\kappa\eta\zeta;\alpha}+
E_{\alpha\beta\mu\nu}E_{\tau\kappa\eta\zeta}
E^{\rho\sigma\gamma\delta}{}_{;\delta}=0=
-6E_{\tau\kappa\eta\zeta;\alpha}
-E_{\alpha\rho\sigma\gamma;\delta}(E_{\tau\kappa\eta\zeta}
E^{\rho\sigma\gamma\delta})=
$$
$$
-6\left(E_{\tau\kappa\eta\zeta;\alpha}+E_{\alpha\tau\kappa\eta;\zeta}+
E_{\zeta\alpha\tau\kappa;
\eta}+E_{\eta\zeta\alpha\tau;\kappa}+E_{\kappa\eta\zeta\alpha;\tau}\right)=0
$$
Or
\begin{equation}\label{compens'}
E_{\alpha\beta\mu\nu;\xi}+E_{\xi\alpha\beta\mu;\nu}+E_{\nu\xi\alpha\beta;
\mu}+E_{\mu\nu\xi\alpha;\beta}+E_{\beta\mu\nu\xi;\alpha}=0
\end{equation}
On the left hand side it is expression to be antisymmetric in five indices $\alpha,\beta,\mu,\nu,
\xi$ in d4-space. The four last terms could play a role of a compensator for
$E_{\alpha\beta\mu\nu;\xi}$.

Next let us consider a pseudovector
$B_\alpha=E_{\alpha\gamma\delta\lambda}
T^{\gamma\delta\lambda}$,
where $T^{\gamma\delta\lambda}$ is an arbitrary (true) antisymmetric in all indices tensor.
Of course it is not a most general representation for the pseudovector.
In other words, the vector $B_\alpha$ is dual to the tensor $T^{\gamma\delta\lambda}$.
As well we can find
$
T^{\rho\gamma\sigma}=-(1/6)B_\alpha E^{\alpha\rho\gamma\sigma}
$

Now let's regard a covariant derivative of a pseudovector $B_\alpha$. For an arbitrary tensor
$T^{\beta\mu\nu}$, by definition is
$$
D_\xi T^{\beta\mu\nu}\equiv T^{\beta\mu\nu}{}_{;\xi}
$$
Then
$$
D_\xi B_\alpha=E_{\alpha\beta\mu\nu}T^{\beta\mu\nu}{}_{;\xi}=B_{\alpha;\xi}-
E_{\alpha\beta\mu\nu;
\xi}T^{\beta\mu\nu}=B_{\alpha;\xi}+\frac{1}{6}E_{\alpha\beta\mu\nu;\xi}
B_\lambda E^{\lambda\beta\mu
\nu}=
$$
$$
=B_{\alpha;\xi}+\frac{1}{6}\gamma'^\lambda_{\alpha\xi}B_\lambda=
B_{\alpha;\xi}-\frac{1}{6}\gamma^\lambda_{\alpha\xi}B_\lambda=
B_{\alpha;\xi}-\gamma_\xi B_\alpha;
$$
Here
$$\gamma'^\lambda_{\alpha\xi}=E_{\alpha\gamma\delta\sigma;\xi}
E^{\lambda\gamma\delta\sigma}
=-E_{\alpha\gamma\delta\sigma}E^{\lambda\gamma\delta\sigma}{}_{;\xi}
=-\gamma^\lambda_{\alpha\xi}
=-6\delta^\lambda_\alpha
\gamma_\xi
$$
So we can see that notion of connection (and covariant derivative) should be extended for the
 pseudotensors
space
\begin{equation}\label{pseconn}
\widetilde{\Gamma}^\lambda_{\alpha\xi}=\Gamma^\lambda_{\alpha\xi}+
\gamma^\lambda_{\alpha\xi}=
\Gamma^\lambda_{\alpha\xi}+\delta^\lambda_\alpha\gamma_\xi
\end{equation}
We would suggest that this connection possesses a torsion, since it is not symmetric in
$\alpha,\xi$. It is not the case in reality. It will be obvious after we calculate a curvature.

The addition to connection $\delta^\lambda_\alpha\gamma_\xi$ is not a tensor, so the
connection~(\ref{pseconn}) transforms otherwise than an ordinary connection. It is a
pseudotensorial connection.

Further we turn to finding curvature in the pseudotensors space. First we find
$$
D_\zeta D_\xi B_\alpha=E_{\alpha\gamma\delta\lambda}
T^{\gamma\delta\lambda}{}_{;\xi;\zeta}=
(E_{\alpha\gamma\delta\lambda}T^{\gamma\delta\lambda}{}_{;\xi})_{;\zeta}-
E_{\alpha\gamma\delta\lambda;\zeta}T^{\gamma\delta\lambda}{}_{;\xi}=
(B_{\alpha;\xi}-
$$
$$
-E_{\alpha\gamma\delta\lambda;\xi}
T^{\gamma\delta\lambda})_{;\zeta}
-E_{\alpha\gamma\delta\lambda;\zeta}T^{\gamma\delta\lambda}{}_{;\xi}=
B_{\alpha;\xi;\zeta}-
E_{\alpha\gamma\delta\lambda;\xi;\zeta}T^{\gamma\delta\lambda}-
E_{\alpha\gamma\delta\lambda;\xi}T^{\gamma\delta\lambda}{}_{;\zeta}-
$$
$$
-E_{\alpha\gamma\delta\lambda;\zeta}T^{\gamma\delta\lambda}{}_{;\xi}
=B_{\alpha;\xi;\zeta}+\frac{1}{6}(E_{\alpha\gamma\delta\lambda;\xi;\zeta}
E^{\sigma\gamma\delta\lambda})
B_\sigma-
E_{\alpha\gamma\delta\lambda;\xi}T^{\gamma\delta\lambda}{}_{;\zeta}-
E_{\alpha\gamma\delta\lambda;\zeta}T^{\gamma\delta\lambda}{}_{;\xi}
$$
Then alternate in $\xi,\zeta$:
$$
(D_\zeta D_\xi-D_\xi D_\zeta)B_\alpha=(B_{\alpha;\xi;\zeta}-
B_{\alpha;\zeta;\xi})+
\frac{1}{6}(E_{\alpha\gamma\delta\lambda;\xi;\zeta}-
E_{\alpha\gamma\delta\lambda;\zeta;\xi})
E^{\sigma\gamma\delta\lambda}B_\sigma
$$
Note that in alternation the terms which could contribute in torsion disappeared.
(first derivatives of $B_\alpha$).

Thus a torsion is absent in the pseudovectors space.

Now the last expression can be reduced to the form
$$
(D_\zeta D_\xi-D_\xi D_\zeta)B_\alpha=B_\lambda(R^\lambda{}_{\alpha\xi\zeta}
+\frac{1}{6}(
\gamma'^\lambda_{\alpha\xi;\zeta}-\gamma'^\lambda_{\alpha\zeta;\xi}))=
B_\lambda\widetilde{R}^\lambda
{}_{\alpha\xi\zeta}=
$$
$$
=B_\lambda(R^\lambda{}_{\alpha\xi\zeta}-
\delta^\lambda_\alpha(\gamma_{\xi,\zeta}-\gamma_{\zeta,\xi}))
$$
On the left hand side we have a pseudotensor, hence, on the right is a pseudotensor too.
$B_\lambda$
is the pseudovector, and so $\widetilde{R}^\lambda{}_{\alpha\xi\zeta}$ is the tensor. Further,
$R^\lambda{}_{\alpha\xi\zeta}$ is the tensor, consequently the term
$\gamma_{\xi,\zeta}-\gamma_{\zeta,\xi}$ is the tensor!

This fact is very important since $\gamma_{\xi}$ is not a vector (noncovariant value).
Moreover $\gamma_{\xi,\zeta}$ is not a tensor. But the difference $\gamma_{\xi,\zeta}-
\gamma_{\zeta,\xi}$ is the tensor, although to be equal to zero.

So, the curvature tensor  in the pseudotensors space is of the form
\begin{equation}\label{pvec_r}
\widetilde{R}^\lambda{}_{\alpha\xi\zeta} =R^\lambda{}_{\alpha\xi\zeta}-
\delta^\lambda_\alpha(\gamma_{\xi,\zeta}-\gamma_{\zeta,\xi})
\end{equation}

We can see that addition to curvature  looks like Weylean one.  On the surface it has
 a number Weylean properties. Still the connection~(\ref{pseconn}) is not invariant under
 weylean rescaling.
For example,  it is scale invariant (for does not depend on metric at all). As it mentioned yet,
$\gamma_\xi$ was not a vector.  There is a vector
(the dilatation generator) in  Weylean curvature instead of $\gamma_\xi$.   
In the given coordinate system this extra curvature term is equal to zero since depends on the
Levi-Civita density and its derivatives. Perhaps that is strangest and unusual property of this
addition. Maybe it is the payment for in what way the Levi-Civita density is defined. In the end,
it is the property
that leads to appearance of the phantom (addition to be equal to zero) in the curvature, meaning of which is unclear.

One can conclude from all that there are two independent characteristic of our
space-time
--- metric $g_{\mu\nu}$ and $\varepsilon_{\alpha\beta\mu\nu}$, so that a full connection are
expressed
in terms of derivatives of these values.

Naturally these zero additions to curvature nothing change in the classic theory. However it is
quite
obvious that the vacuum structure may essentially change in  nonexisting as yet quantum gravity.
May be these additions improve the situation with quantizing gravity.
Topological structure of vacuum seems to get  more interesting.

Here are analogous results for dual (pseudo)tensor of the second rank
${*\!F}_{\alpha\beta}=1/2E_{\alpha\beta\gamma\delta}F^{\gamma\delta}$
and pseudotensor of the third rank
 $P_{\alpha\beta\gamma}=E_{\alpha\beta\gamma\delta}Q^\delta$:
$$
(D_{\zeta}D_{\xi}-D_{\xi}D_{\zeta}){*\!F}_{\alpha\beta}=
{*\!F}_{\alpha\lambda}\left(R^\lambda{}_{\beta
\xi\zeta}-\frac{1}{2}\delta^\lambda_\beta(\gamma_{\xi,\zeta}-
\gamma_{\zeta,\xi})\right)+
$$
$$
{*\!F}_{\lambda\beta}\left(R^\lambda{}_{\alpha
\xi\zeta}-\frac{1}{2}\delta^\lambda_\alpha(\gamma_{\xi,\zeta}-
\gamma_{\zeta,\xi})\right)
$$
Deduce from that
\begin{equation}\label{pf_r}
\widetilde{R}^\lambda{}_{\alpha\xi\zeta} =R^\lambda{}_{\alpha\xi\zeta}-
\frac{1}{2}\delta^\lambda_\alpha(\gamma_{\xi,\zeta}-\gamma_{\zeta,\xi})
\end{equation}
This does not coincide with~(\ref{pvec_r}).

Alternated second derivative of the pseudotensor $P_{\alpha\beta\gamma}$
is of the form
$$
(D_{\zeta}D_{\xi}-D_{\xi}D_{\zeta})P_{\alpha\beta\mu}=P_{\alpha\beta\lambda}
R^\lambda{}_{\mu\xi
\zeta}+P_{\alpha\lambda\mu}R^\lambda{}_{\beta\xi\zeta}+P_{\lambda\beta\mu}
R^\lambda{}_{\alpha\xi
\zeta}-P_{\alpha\beta\mu}(\gamma_{\xi,\zeta}-\gamma_{\zeta,\xi})
$$
And the curvature tensor in the third rank pseudotensors has the form
\begin{equation}\label{pp_r}
\widetilde{R}^\lambda{}_{\alpha\xi\zeta}=R^\lambda{}_{\alpha\xi\zeta}-
\frac{1}{3}
\delta^\lambda_\alpha(\gamma_{\xi,\zeta}-\gamma_{\zeta,\xi}),
\end{equation}
that again  does not coincide either with~(\ref{pvec_r}), or with~(\ref{pf_r}).
 Thus the form of the Weylean addition to curvature in the pseudotensorial space depends on
 the pseudotensor rank. There does not exist universal curvature tensor in such a space.

\section{A torsioned connection example  constructed from metric and Levi-Civita tensor}

In the previous section~(\ref{Qpreob}) there has been constructed the vector $G_{\xi}(q_0)$.
This make it possible to build connections which are distinguished from the metric those. Let us
consider
a geometry example with the extended connection~(\ref{pconnect})
$$
\widetilde{\Gamma}_{\alpha\mu\xi}=\Gamma_{\alpha\mu\xi}+G^\lambda(q_0)E_{\lambda\alpha\mu\xi}=
\Gamma_{\alpha\mu\xi}+\gamma_{\alpha\mu\xi},\quad \gamma_{\alpha\mu\xi}=\gamma_{[\alpha\mu\xi]}=
G^\lambda(q_0)E_{\lambda\alpha\mu\xi}
$$
This connection holds metricity condition, differs from Riemannian connection by a pseudotensor
and leads to a combined curvature tensor (i.e. tensor plus pseudotensor).

In order to obtain a curvature tensor let's calculate the alternated second covariant derivative
of the arbitrary true tensor $A_{\alpha}$. In accordance with~(\ref{pconnect})
\begin{equation}\label{covder}
D_\xi A_\alpha=A_{\alpha;\xi}-\gamma^\lambda{}_{\alpha\xi}A_\lambda
\end{equation}
So the covariant derivative of the true vector is a combined tensor.

We shall need the Levi-Civita tensor covariant derivative as well. Taking into
account~(\ref{compens}) and the
covariant
differentiation rule, it must be determined as follows
$$
D_\xi E_{\alpha\beta\mu\nu}=(E_{\alpha\beta\mu\nu;\xi}-\frac{1}{24}\gamma_\xi E_{\alpha\beta\mu
\nu})-
\gamma^\lambda{}_{\alpha\xi}E_{\lambda\beta\mu\nu}-
\gamma^\lambda{}_{\beta\xi}E_{\alpha\lambda\mu\nu}-
\gamma^\lambda{}_{\mu\xi}E_{\alpha\beta\lambda\nu}-
$$
$$
-\gamma^\lambda{}_{\nu\xi}E_{\alpha\beta\mu\lambda}
=\gamma^\lambda{}_{\lambda\xi}E_{\alpha\beta\mu\nu}=0
$$
So
$$
D_\xi E_{\alpha\beta\mu\nu}=0
$$
Further calculate taking into account~(\ref{covder})
$$
D_\zeta D_\xi A_\alpha=D_\zeta A_{\alpha;\xi}-D_\zeta (\gamma^\lambda{}_{\alpha\xi}A_\lambda)=
A_{\alpha;\xi;\zeta}-
\gamma^\lambda{}_{\alpha\xi}A_{\lambda;\xi}-\gamma^\lambda{}_{\xi\zeta}A_{\alpha;\lambda}-
$$
$$
-\gamma^\lambda{}_{\alpha\xi}D_\zeta A_\lambda-A_\lambda D_\zeta\gamma^\lambda{}_{\alpha\xi}
$$
Then alternate in $\xi,\zeta$
$$
(D_\zeta D_\xi-D_\xi D_\zeta)A_\alpha=A_\lambda\left(R^\lambda{}_{\alpha\xi\zeta}+
(\gamma^\delta{}_{\alpha\xi}\gamma^\lambda{}_{\delta\zeta}-
\gamma^\delta{}_{\alpha\zeta}\gamma^\lambda{}_{\delta\xi}\right.)+
$$
$$
+\left.(D_\xi\gamma^\lambda{}_{\alpha\zeta}-
D_\zeta\gamma^\lambda{}_{\alpha\xi})\right)-2A_{\alpha;\lambda}\gamma^\lambda{}_{\xi\zeta}
$$
The new curvature tensor $\widetilde{R}^\lambda{}_{\alpha\xi\zeta}$ is of the form
\begin{equation}\label{tcurv}
\widetilde{R}^\lambda{}_{\alpha\xi\zeta}=
R^\lambda{}_{\alpha\xi\zeta}+
(\gamma^\delta{}_{\alpha\xi}\gamma^\lambda{}_{\delta\zeta}-
\gamma^\delta{}_{\alpha\zeta}\gamma^\lambda{}_{\delta\xi})+
(D_\xi\gamma^\lambda{}_{\alpha\zeta}-
D_\zeta\gamma^\lambda{}_{\alpha\xi})
\end{equation}
This includes three terms: a Riemann-Christoffel tensor, plus the tensorial and pseudotensorial
additions due to the extended connection. Thus we have a combined curvature tensor.

It is not difficult to obtain contracted curvature tensors. First contract~(\ref{tcurv}) in
$\lambda,\xi$:
$$
\widetilde{R}_{\alpha\zeta}=\widetilde{R}^\lambda{}_{\alpha\lambda\zeta}=R_{\alpha\zeta}+
\gamma^\delta{}_{\alpha\lambda}\gamma^\lambda{}_{\delta\zeta}+D_\lambda\gamma^\lambda{}_{\alpha\zeta}
$$
Here is a useful formula for an inquiry
$$
D_\xi\gamma^\lambda{}_{\alpha\zeta}-D_\zeta\gamma^\lambda{}_{\alpha\xi}=g^{\lambda\sigma}
(G^\delta_{;\xi}E_{\delta\sigma\alpha\zeta}-G^\delta_{;\zeta}E_{\delta\sigma\alpha\xi})
$$
Now one obtains
\begin{equation}\label{ricci}
\widetilde{R}_{\alpha\zeta}=R_{\alpha\zeta}+
\gamma^\delta{}_{\alpha\lambda}\gamma^\lambda{}_{\delta\zeta}+g^{\lambda\sigma}
G^\delta_{;\lambda}E_{\delta\sigma\alpha\zeta},\,\gamma^\delta{}_{\alpha\lambda}\gamma^\lambda{}_{\delta\zeta}
=2(G_\alpha G_\zeta-G^\lambda G_\lambda g_{\alpha\zeta})
\end{equation}
So the tensor $\widetilde{R}_{\alpha\zeta}$ contains an antisymmetric part to be a pseudotensor.
However, the part is equal to zero because of the vector $G_\mu$ consists of two terms, the first
of which is  a gradient and the second one is a phantom. It is the antisymmetric part that is the
phantom
$$
\widetilde{R}_{[\alpha\zeta]}=-\frac{1}{48}(\gamma^{\delta,\lambda}-
\gamma^{\lambda,\delta})E_{\delta\lambda\alpha\zeta}
$$

One more contracting, obtain the scalar curvature
\begin{equation}\label{skalr}
\widetilde{R}=R+\gamma^{\rho\lambda\sigma}\gamma_{\rho\lambda\sigma}=
R-6G^\lambda G_\lambda
\end{equation}
Note additions to the curvature compared to Riemannian case. These are small in weak fields
since they include
the metric derivatives. So, Einstein's equations will include the small corrections as well.

It is of interest to calculate a twice dual tensor ${*\!\widetilde{R}\!*}_{\alpha\beta\mu\nu}$~\cite{ko}.
$$
{{*\!\widetilde{R}\!*}_{\alpha\beta\mu\nu}}=-\widetilde{R}_{\mu\nu\alpha\beta}+
g_{\mu\alpha}\widetilde{R}_{\nu\beta}+g_{\nu\beta}\widetilde{R}_{\mu\alpha}-
g_{\mu\beta}\widetilde{R}_{\nu\alpha}-g_{\nu\alpha}\widetilde{R}_{\mu\beta}-
\frac{1}{2}\widetilde{R}g_{\alpha\beta\mu\nu}
$$

The Bianchi identity for ~(\ref{tcurv}) can be simply proved by  going to local Lorentz system.

\section{Curvature in the combined tensors space}

A sum of a tensor and a pseudotensor of the same rank we shall call
a {\em combined} tensor. In the Introduction there were given motivations of the treatment. 
We will consider a special case of the combined tensor
$$
C_{\alpha\beta}=aF_{\alpha\beta}+b{*\!F}_{\alpha\beta},
$$
where $F_{\alpha\beta}$ is an arbitrary antisymmetric tensor, and $a,b$ are arbitrary constants.
We are interested in curvature in the $C_{\alpha\beta}$ tensors space. In the long run we'll fail
to find any curvature tensor of the fourth rank in the space us to be interested in. However,
it is possible to obtain some the 6-rank tensor which represents
the curvature measure in the combined tensors space.

First of all write the tensor as follows
$$
C_{\alpha\beta}=aF_{\alpha\beta}+b{*\!F}_{\alpha\beta}=
\frac{1}{2}(ag_{\alpha\beta\gamma\delta}+bE_{\alpha\beta\gamma\delta})
F^{\gamma\delta}=d_{\alpha\beta\gamma\delta}F^{\gamma\delta},
$$
where
$$
d_{\alpha\beta\gamma\delta}=
\frac{1}{2}(ag_{\alpha\beta\gamma\delta}+bE_{\alpha\beta\gamma\delta}),\quad
g_{\alpha\beta\gamma\delta}=g_{\alpha\gamma}g_{\beta\delta}-g_{\alpha\delta}g_{\beta\gamma}
$$
Then introduce
$$
\tilde{d}^{\mu\nu\rho\sigma}=\frac{1}{a^2+b^2}(ag^{\mu\nu\rho\sigma}-bE^{\mu\nu\rho\sigma})
$$
so that
$$
d_{\alpha\beta\gamma\delta}\tilde{d}^{\alpha\beta\lambda\sigma}=\delta^{\lambda\sigma}_{\gamma\delta}
$$
And let's write two more identities
\begin{equation}\label{tozhd}
F^{\lambda\sigma}=\frac{1}{2}\delta^{\lambda\sigma}_{\gamma\delta}F^{\gamma\delta}=\frac{1}{2}
d_{\alpha\beta\gamma\delta}\tilde{d}^{\alpha\beta\lambda\sigma}F^{\gamma\delta}=
\frac{1}{2}C_{\alpha\beta}\tilde{d}^{\alpha\beta\lambda\sigma}
\end{equation}

According to technic worked out above, one can  obtain  the second covariant
derivatives commutator of the combined tensor
$C_{\alpha\beta}$
$$
(D_\zeta D_\xi-D_\xi D_\zeta)C_{\alpha\beta}=a(F_{\alpha\lambda}R^\lambda{}_{\beta\xi\zeta}+
F_{\lambda\beta}R^\lambda{}_{\alpha\xi\zeta})+
b({*\!F}_{\alpha\lambda}\widetilde{R}^\lambda{}_{\beta\xi\zeta}+
{*\!F}_{\lambda\beta}\widetilde{R}^\lambda{}_{\alpha\xi\zeta})
$$
Here
 $\widetilde{R}^\lambda{}_{\alpha\xi\zeta}$ is given by~(\ref{pf_r}).

Consider an expression
 $aF_{\alpha\lambda}R^\lambda{}_{\beta\xi\zeta}+b{*\!F}_{\alpha\lambda}\widetilde{R}^
\lambda{}_{\beta\xi\zeta}$.
The task is to rewrite this in terms of the combined tensor
$C_{\alpha\lambda}$. Then the coefficient at the term could be treated as the curvature measure
in the combined tensors space. Using~(\ref{tozhd}), the expression can be transformed to the form
$$
aF_{\alpha\lambda}R^\lambda{}_{\beta\xi\zeta}+b{*\!F}_{\alpha\lambda}\widetilde{R}^
\lambda{}_{\beta\xi\zeta}
=\frac{1}{2}F^{\gamma\delta}(ag_{\alpha\lambda\gamma\delta}R^
\lambda{}_{\beta\xi\zeta}+bE_{\alpha\lambda\gamma\delta}\widetilde{R}^\lambda{}_{\beta\xi\zeta})=
$$
$$
\frac{1}{4}C_{\rho\sigma}\tilde{d}^{\rho\sigma\gamma\delta}(ag_{\alpha\lambda\gamma\delta}
R^\lambda{}_{\beta\xi\zeta}+bE_{\alpha\lambda\gamma\delta}\widetilde{R}^\lambda{}_{\beta\xi\zeta})
$$
Unfortunately, on the right hand side we have $C_{\rho\sigma}$ with the dummy subscripts
rather than $C_{\alpha\lambda}$. Therefore as the curvature measure one has to take the 6-rank
combined tensor:
$$
\mathcal{R}^{\rho\sigma}{}_{\alpha\beta\xi\zeta}=
\tilde{d}^{\rho\sigma\gamma\delta}(ag_{\alpha\lambda\gamma\delta}
R^\lambda{}_{\beta\xi\zeta}+bE_{\alpha\lambda\gamma\delta}\widetilde{R}^\lambda{}_{\beta\xi\zeta})=
$$
$$
\frac{1}{a^2+b^2}(ag^{\mu\nu\rho\sigma}-bE^{\nu\rho\sigma})
(ag_{\alpha\lambda\gamma\delta}
R^\lambda{}_{\beta\xi\zeta}+bE_{\alpha\lambda\gamma\delta}\widetilde{R}^\lambda{}_{\beta\xi\zeta})=
$$
$$
=\frac{2}{a^2+b^2}\left(aR^\lambda{}_{\beta\xi\zeta}(a\delta^{\rho\sigma}_{\alpha\lambda}-
bE^{\rho\sigma}{}_{\alpha\lambda})+b\widetilde{R}^\lambda{}_{\beta\xi\zeta}(b\delta^{\rho\sigma}_
{\alpha\lambda}+aE^{\rho\sigma}{}_{\alpha\lambda})\right)
$$

\section{Conclusion}

So, the correct covariant derivative of the Levi-Civita tensor has been constructed. First, this
 implies
 appearance of the phantom terms in the curvature tensor that interpretation is unclear. Second,
 it permits to build the vector $G_\mu$. Availability of the vector opens opportunities for
 extension of connection and construction of various curvature tensors. It is shown that the
 curvature
 tensor can contain the pseudotensor terms. For Yang-Mills fields, similar result has been obtained
 in~\cite{kosh} and that led to analytical continuation of instantons. It would be desirable to
  do something like that in gravity.

 These curvature tensors have more general properties and
 symmetries. New approach to deduction of the gravity equations based on the duality properties
 of the curvature tensor
 has been proposed in~\cite{ko}.
 As a development of the method in publication to follow, there will be obtained equations of the
 form (besides Einstein's)
 $$
\widetilde{R}_{[\alpha\mu]}=T_{[\alpha\mu]}
$$
where $T_{[\alpha\mu]}$- is an antisymmetric part of the energy-momentum tensor, and
$\widetilde{R}_{[\alpha\mu]}$ is an antisymmetric part of the Ricci tensor.

  The fact that the curvature tensor includes phantoms is unusual and difficult for interpretation.
   That may be useful
 in quantizing  gravity.

\end{document}